# Atomic scale investigation of Cr precipitation in copper.

A. Chbihi[1], X. Sauvage[1]*, D. Blavette[1,2]

1- University of Rouen, GPM, UMR CNRS 6634, BP 12, Avenue de l'Université, 76801 Saint-Etienne du Rouvray, France

2- Institut Universitaire de France

* Corresponding author: xavier.sauvage@univ-rouen.fr

**Abstract**

The early stage of the chromium precipitation in copper was analyzed at the atomic scale by Atom Probe Tomography (APT). Quantitative data about the precipitate size, 3D shape, density, composition and volume fraction were obtained in a Cu-1Cr-0.1Zr (wt.%) commercial alloy aged at 713K. Surprisingly, nanoscaled precipitates exhibit various shapes (spherical, plates and ellipsoid) and contain a large amount of Cu (up to 50%), in contradiction with the equilibrium Cu-Cr phase diagram. APT data also show that some impurities (Fe) may segregate along Cu/Cr interfaces. The concomitant evolution of the precipitate shape and composition as a function of the aging time is discussed. A special emphasis is given on the competition between interfacial and elastic energy and on the role of Fe segregation.

# 1. Introduction

Age hardenable Cu-Cr alloys and their minor modifications (including for example Cu-Cr-Zr alloys) are used in numerous applications where an excellent combination of mechanical strength and electrical or thermal conductivity is required. Thus, they are attractive candidates for railway contact wires [1], electrodes for spot welding, heat exchangers and more recently they are being considered for the design of plasma facing components and divertor plates of fusion devices [2-4]. The high strength (comparing to pure copper) is achieved thanks to typical precipitations treatment that give rise to the nucleation of a high density of Cr nanoscaled precipitates [1, 3-7]. The high conductivity results from the extremely low solubility of Cr in fcc Cu at temperatures below 500°C [8].

Dilute Cu-Cr alloys have been investigated since more than a century with the first microscopic study reported by Guillet in 1906 [9] and the first data on the binary phase diagram reported by Hindrichs in 1908 [10]. The precipitation kinetic was first investigated using indirect methods like X-ray diffraction or dilatometry [5, 11]. These experiments raised some fundamental questions about the shape and the structure of the precipitates. In the 70's and later, nanoscaled Cr precipitates could be observed by Transmission Electron Microscopy (TEM) [5, 7, 12-14], starting a long controversy on their crystallographic structure and orientation relationships (OR) with the fcc copper matrix. The equilibrium crystallographic structure of Cr is bcc, but in the early stage of precipitation it has been proposed by several authors that they could exhibit a metastable fcc structure [2, 7, 14]. Part of the enigma was finally solved by Fujii and co-authors 30 years later thanks to High Resolution TEM (HRTEM) observations [15]. They have shown that even if some typical lobe-lobe contrast usually attributed to coherent iso-structural precipitates could be observed, they are all bcc in structure even in the early stage of precipitation. They also demonstrated that two OR may exist (known as the Nishiyama-Wassermann (NW) and the Kurdjumov-Sachs (KS)) but for longer aging time, the particles exhibiting a KS-OR grow at the expense of those having a NW-OR.

In any case, nanoscaled Cr precipitates induce large lattice distortions that make them difficult to observe by TEM. Thus, there are still some ambiguities about their exact shape and size, especially in the early stage of nucleation. Some authors describe them as GP zones (i.e. few atomic planes) [16], others reporting about spherical [3], ellipsoid [15], plate [5, 12], rod shaped precipitates [12, 13] or even hexagonal [17]. But very recently, Hatakeyama and co-authors have used Atom Probe Tomography (APT) to analyze such precipitates [18, 19]. They have shown that this technique can provide some reliable information about the size, the shape, the composition and the density of the Cr precipitates. Unfortunately they mostly focussed their work on large precipitates resulting from coarsening that occurs during re-aging treatment at 600°C.

During the past ten years, modified Cu-Cr alloys and especially those from the ternary Cu-Cr-Zr system have been widely investigated because of their improved combination of mechanical and electrical (or thermal) properties [1-4, 6, 18-20]. The addition of Mg or Fe has also been proposed by some authors [17, 21-22]. In such more complex alloys, other phases were observed like $Cu_4Zr$ or $CrCu_2Zr$ [3, 16, 22] and some Zr and Fe segregations along Cu/Cr interfaces were also reported [18]. Moreover, the precipitation kinetic of Cr particles seems modified since Batra and co-authors identified only fcc Cr particles with a cube on cube OR and also a metastable ordered phase with a possible B2 structure [20].

The nucleation of precipitates with a crystallographic structure different from that of the parent matrix is a fundamental problem that has been addressed for a long time by many authors. Geometrical criteria using the invariant-line concept have been successfully adopted to explain ORs between precipitates and matrices [23]. However, one should note that in Cu based systems very different scenario may happen in the early stage of decomposition depending on the solute element. This may indicate that thermodynamics, elastic anisotropy and interfacial energy anisotropy play also a critical role in the precipitation kinetic. For example, in dilute Cu-Be alloy, a complex sequence of metastable phases starting from GP zones has been reported before the formation of the stable B2 phase [24]. In the Cu-Fe system, the coherent metastable fcc γ-Fe particles that nucleate first can be stable up to 50 nm in size [25, 26], while in the symmetric system (dilutes Fe-Cu alloys) copper particles nucleate as bcc but start to progressively transform into the stable fcc structure when they are only few nanometer in size [27]. For this later system, it is also interesting to note that in the early stage, precipitates contain a significant amount of Fe, much higher than the solubility limit given by the phase diagram [28, 29]. For the Cu-Cr system, as already discussed, it has been proposed that the phase separation involves the nucleation of the Cr fcc metastable phase, and even a long range ordered structure with an equi-atomic composition has been proposed (namely a B2 phase [20]). Thus, the aim of this work was to get a clear picture of nanoscaled Cr precipitates in the early stages of nucleation using a combination of TEM and APT. This later technique can provide an accurate description of the 3D shape, the volume, the distribution, the density, the volume fraction and the composition of precipitates.

For this study, a Cu-Cr-Zr alloy was chosen since this kind of alloy has attracted a lot of interest during the past decade. Different kinds of precipitate were exhibited and a special emphasis was given on the existing relationships between the size, the morphology and the composition of the precipitates for a better understanding of the precipitation kinetic.

**2. Experimental**

The alloy investigated in the present study is a commercial Cu-1Cr-0.1Zr alloy (wt.%) received in the peak-aged hardening state. The material was homogenized at 1323 K for 1 h in Ar atmosphere followed by water quenching and isothermal aging at 713 K. This temperature is slightly lower than conventional treatments (typically in a range of 733 to 753 K [1-4]) to slow down the precipitation kinetic and to really get the early stage of precipitation. The age hardening behaviour under such condition is reported in Fig. 1, showing that a saturation of the hardness is reached after 10h (155 HV ± 8), in good agreement with published data on similar alloys [5, 6]. The Vickers micro hardness of the alloy was measured with a BUEHLER Micromet 2003 machine, using a 300 g load (2.94 N), each plotted value being the average of ten measurements.

As expected, the increase of the hardness is followed by a concomitant increase of the conductivity, up to 76 ± 10 % IACS (see table 1), in good agreement with published data on similar alloys [5, 17]. Resistivity measurements were performed by the four-point method using a micro ohmmeter OM21 of AOIP® (Resolution of 0.1 μΩ, precision of 0.03 %). Provided data are the average of at least 20 measurements.

Precipitates were characterized by conventional TEM using a JEOL 2000FX and by analytical Scanning TEM (STEM) using a JEOL 2100F, both microscopes operating at 200kV. Energy Dispersive X-ray Spectroscopy (EDX) was carried out with a JEOL JED-2300T system. Electron Energy Loss Spectroscopy (EELS) was performed using a Gatan ENFINA spectrometer. TEM samples were prepared by jet-polishing (TUNEPOL 3 - Struers) at 243K, with a voltage of 15V and using a mixture of 20 % nitric acid and 80 % methanol (in volume) as electrolyte.

APT analyses were performed using a CAMECA Laser Wide Angle Tomography Atom Probe (LAWATAP) [30]. Specimens were prepared using standard electro-polishing methods (10V, electrolyte: $H_3PO_4$ + 30% $H_2O$,) and field evaporated in ultra-high vacuum conditions at 20 K with femtosecond laser pulses (pulse repetition rate of 100 kHz and wavelength of 515 nm).

In the as quenched state, APT data revealed that during the solution treatment up to 0.78 ± 0.09 at.% Cr atoms are dissolved in the fcc copper matrix (table 1), in good agreement with the equilibrium phase diagram [8]. Although it was not mentioned in the nominal composition of the commercial alloy, a significant amount of Fe in solid solution was also detected (0.09 ± 0.01 at.%). It is important to note that these two solute elements are homogeneously distributed in the solutionized alloy (APT 3D reconstructed volume not shown here), indicating that any clustering or GP zone formation did not occur during quenching as suggested by some authors [5, 21]. The amount of Zr in solid solution after the homogenization treatment was always below the detection limit of the atom probe (i.e. 100 ppm). Large precipitates (i.e. 0.3 to 1 μm in diameter) containing a significant amount of Zr do exist in the as received alloy and were observed by TEM and analyzed by EDX (data not shown here). However, such Cu-Zr intermetallic phases are stable up to very high temperatures [31] and as already reported by Apello and Fenici [32], they probably do not dissolve during the short homogenization treatment.

It is well known that APT 3D reconstructed volumes can be affected by local magnification effects if different phases with different evaporation fields are evaporated simultaneously [33]. The predicted evaporation fields for chromium ($Cr^+$, 27 $V.nm^{-1}$) and copper ($Cu^+$ 30 $V.nm^{-1}$) are slightly different [34]. However, during the present study and under the selected APT analysis conditions, chromium (respectively copper) ions were almost exclusively collected as $Cr^{2+}$ (respectively $Cu^+$). The theoretical evaporation field of $Cr^{2+}$ (29 $V.nm^{-1}$ [34]) is much closer to that of $Cu^+$, thus non significant local magnification effects are expected. To check this point, Field Ion Microscopy was performed on the alloy in the peak-aged hardening state with a high density of Cr precipitates (Fig. 2). Only the typical ring contrast of the Cu fcc crystallographic poles could be observed, even upon the evaporation of a large volume. These data clearly indicate that local magnification effects, if any, would not significantly affect 3D reconstructions and especially precipitate compositions or concentration gradients.

# 3. Microscopic features of Cr precipitates nucleated in the fcc Cu matrix

The copper alloy aged during 10h at 713K was first investigated by TEM. Cr precipitates that have nucleated during the aging treatment are very small, typically few nanometers and their orientation relationships with the fcc copper matrix usually induce strong local distortions [23]. This leads to some complex contrasts on bright field TEM images which make the estimation of the precipitate density, morphology, distribution and size complicated. However, a careful observation of the bright field image obtained near the (002) zone axis (Fig. 3 (c)) revealed three kinds of contrasts associated to these precipitates. First, some typical coffee-bean contrasts (Fig. 3 (d)) with an extinction line along the <001>Cu,fcc direction and corresponding to coherent fcc precipitates [7, 14]. Then, some Moiré contrasts (Fig. 3(e)) with fringes aligned along a direction close to <110>Cu,fcc typical of bcc Cr precipitates exhibiting a NW-OR or KS-OR [15]. Last, dark precipitates (Fig. 3 (f)) elongated along a direction close to <110>Cu,fcc which might be attributed to larger bcc Cr precipitates. The Selected Area Electron Diffraction (SAED) pattern (Fig. 3 (a)) obtained in the (002) zone axis provides further information. For more clarity, spots are indexed in the Fig. 3 (b). First it is interesting to note that all fcc Cu spots are streaked along the <110> direction which is attributed to significant distortions of the crystal lattice. Then, there are some weak spots that can undoubtedly be attributed to the scattering of electrons by (002)Cr,bcc atomic planes. A careful observation reveals that there are three sets of these spots tilted respectively by 0°, 30° and 60° from the <110> direction of the fcc matrix. Such features are consistent with Cr precipitates with a bcc structure having a NW-OR typical of the early stage of precipitation [15]. Last, some additional weak spots also appear close to the transmitted beam. They might be attributed to double diffraction or to (001)Cr,bcc superlattice reflections of an ordered bcc phase as reported by Batra and co-authors [20]. This latter possibility would be discussed in the next section.

From these TEM data collected after 10h of aging at 713K, it seems that three kinds of precipitates co-exist within the microstructure: some of them are most probably fcc and coherent with the copper matrix (coffee-bean contrast, 20% ± 5), others are most probably bcc whether with a NW-OR or a KS-OR (Moiré fringes contrast, 25% ± 5, dark contrast and elongated, 55% ± 5). However, there are obviously some data missing to get a full picture of the precipitation kinetic and among them the size, the composition, the morphology or the density.

Part of this information was collected using analytical TEM. Using EDX mapping in the STEM mode the high density of nanoscaled chromium precipitates was clearly revealed (Fig.

4 (a)). Over the 120 x 120 nm$^2$ area there are about 40 precipitates, thus assuming that the foil thickness was in a range of 20 to 50 nm, it leads to a precipitate density in a range of 0.5 to 1.5 10$^{23}$ m$^{-3}$. However, one should note that due to the low spatial resolution and the background noise some of the smallest precipitates are probably not detected leading to an under-estimation of the overall density. High spatial resolution EELS mapping was also performed. Three chromium precipitates, about 5 nm in size, are clearly revealed (Fig. 4 (b, c)). Two of them appear roughly spherical, while the third one exhibits an ellipsoid morphology with a length over width ratio of about two. This later looks similar to the elongated precipitate revealed on the bright field TEM image (Fig. 3 (f)). However, these images provide only a 2D projection of precipitates, thus one can only speculate about the three-dimensional morphology. Moreover, it is impossible to measure accurately their composition because they are fully embedded in the copper matrix. Therefore APT analyses were performed.

The alloy was analyzed after 5 and 10h of aging at 713 K and the nanoscaled Cr rich precipitates are clearly exhibited in the 3D reconstructed volumes (Fig. 5). A quick inspection of these images immediately shows that after 10h of aging the precipitate density is lower while the mean precipitate size is larger. Filtering the data with a Cr concentration threshold of 5 at.% allowed the identification of all precipitates and thus the estimation of the precipitate density (see table 1). The measured density for 10h of aging is twice higher than the estimation performed on the basis of analytical TEM. This might be attributed to the higher capability of APT for the detection of the smallest precipitates. A similar filtering procedure was used to remove atoms linked to precipitates in analyzed volumes and to estimate the amount of Cr in solid solution in the matrix. It is interesting to note that even if the precipitation of chromium is not completed after 5h (0.14at.%Cr left in solid solution versus 0.02at.% after 10h, see table 1), the coarsening regime has already started. As expected, the decrease of the Cr concentration in solid solution in the copper matrix leads to a significant increase of the conductivity (table 1). To the knowledge of the authors, there is no data available in the literature about the Cr solubility limit at a temperature as low as 713 K. However, according to Nagata and Nishikawa [5] it should be less than 0.05 at.%. Thus, one may reasonably assume that after 10h of aging at 713K the Cr concentration in the matrix has reached the thermodynamical equilibrium.

The 3D reconstructed volumes displayed in the Fig. 5 also clearly show that the precipitate morphology significantly changes between 5 and 10h of aging. Indeed, in the early stage of precipitation (5h), they look more spherical. In fact, three kinds of precipitates with different shapes could be found in these two volumes: spheres (S), ellipsoids (E) and plates (P). The respective abundance and the size distribution of these different precipitates are provided in the Fig. 6 revealing three important features: (i) for all kinds of precipitates, size distributions are large, (ii) the smallest precipitates are spherical while the largest are ellipsoid or plate shaped, (iii) the proportion of spherical precipitates dramatically decreases with the aging time (from 26 to 6 %) while the number of plates strongly increases (from 37 to 64%). One should note that three kinds of precipitates were also observed by TEM, thus these different shapes could be related to different crystallographic structures (fcc or bcc) and / or different OR (KS or NW). No systematic orientation relatively to each others could be found for plate or ellipsoid shaped precipitates in 3D reconstructed volumes. This might be due to the large number of variants offered by both the NW-OR and the KS-OR.

The composition in each precipitate was determined using concentration profiles with small sampling volumes (typical thickness 1 nm). Representative profiles related to spherical, ellipsoid and plate shaped precipitates are displayed in the Fig. 7. The 1nm large Cr gradient appearing at the matrix/precipitate interface can be exclusively related to the thickness of the sampling volume, thus interfaces are chemically sharp. However, all these precipitates contain a significant amount of Cu which is surprisingly much higher than the theoretical prediction of the Cu-Cr binary phase diagram [8]. Local magnification effects were ruled out in the previous section, thus this feature cannot be attributed to analysis artefact. The chromium concentration is only about 50at.% in spherical precipitates (Fig. 7 (a)), up to about 80at.% in ellipsoids (Fig. 7 (b)) and up to about 90% in plates (Fig. 7 (c)).

Spherical precipitates are the smallest, so it seems that smallest precipitates contain a lowest proportion of Cr. Considering the large size distribution of all kinds of precipitates (Fig. 6), the Cr concentration was plotted as a function of the precipitate size for all precipitates and for 5 and 10h of aging (Fig. 8). The trend is confirmed and appears especially clear for the material aged 5h where the mean precipitate volume is smaller (Fig. 8 (a)): the smaller the precipitates the lower their Cr content.

It is interesting to note that, additionally to Cr, precipitates also contain a significant amount of Fe (Fig. 7). For the smallest precipitates (i.e. spheres), Fe atoms seem homogeneously mixed with Cr (Fig. 7 (a)). But for larger precipitates and especially plate shaped, Fe atoms are segregated along the Cu/Cr interface (Fig. 7 (c)). These observations are in agreement with recently published data from Hatakeyama and co-authors [18]. However, one should note that in the present study, Fe atoms were not found homogeneously distributed on the precipitate surface and there is not a full coverage of the interface.

## 4. Discussion about the nucleation and growth of precipitates

Bright field TEM investigations of the early stages of precipitation have revealed the presence of both coherent fcc precipitates as well as bcc Cr-enriched precipitates with two crystal orientation relationships (NW and KS), The proportion of each kind of precipitate is difficult to estimate from this only TEM information. Only HR-TEM of individual precipitates could provide precise information on their OR (NW or KS). The precipitate density was estimated thanks to analytical TEM. APT investigations revealed three morphologies: spherical, ellipsoid and plate shape. Fujii and co-authors have shown that bcc Cr precipitates appear elongated in 2D HR-TEM images [15]. It is also interesting to note that for the NW-OR there is a small misfit (2.3%) only along one direction, namely $(111)_{Cu} // (110)_{Cr}$, while for the KS-OR there is a second low misfit direction (2.2%), namely $[-101]_{Cu} / [-111]_{Cr}$. It seems therefore realistic to think that precipitates will favorably grow along these low misfit directions, leading to some ellipsoids for the NW-OR and plate shaped for the KS-OR. Chromium precipitates with an fcc structure and a cube on cube OR with the fcc matrix should of course grow with a spherical shape resulting from the isotropic distortions. Moreover, APT data show that spherical precipitates are much smaller than others and they are more numerous upon the early stage of precipitation, but also that the proportion of plate shaped precipitates dramatically increases with the aging time and they are bigger in average than others. Fujii and co-authors have shown that bcc chromium precipitates with a KS-OR grow at the expense of those having a NW-OR [15]. As a consequence, one may conclude that ellipsoid shaped precipitates are bcc with a NW-OR while plate shaped are bcc with a KS-OR. The precipitation sequence in the dilute Cu-Cr system is therefore thought to be as following: (i) nucleation of coherent fcc spherical precipitates, (ii) growth and transformation of fcc spherical precipitates into the bcc structure, (iii) growth of bcc precipitates that have a KS-OR at the expense of those having a NW-OR. APT data suggest that certain precipitates enter into the coarsening stage while nucleation of new embryos go on (i.e. precipitates have not nucleate at the same time).

APT analyses and TEM observations also strongly suggest that precipitation of the Cr-rich equilibrium bcc phase initiates through the nucleation of fcc precipitates that are coherent with the Cu-rich fcc parent phase. Such a precipitation pathway has been observed in numerous systems, in particular in Al base alloys where coherent GP zones nucleate before the stable phase (e.g. GP zones in AlCu). The physical reason behind this process is well known and is associated with the much lower interfacial energy of coherent interfaces that reduces the nucleation barrier compared to incoherent or semi-coherent interfaces. Experiments show that in the early stage of precipitation most of precipitates contain a significant amount of Cu. Similar features were reported for the precipitation of Cu in bcc-Fe where non equilibrium bcc Cu precipitates nucleate with a significant content of Fe [28, 29]. Present researches reveal that in CuCr the transient metastable fcc phase have a Cr content close to 50 at.%. This is rather surprising if we consider the very high enthalpy of mixing of Cu with Cr in the fcc structure for such a composition ($\Delta H_m$ close to 20 kJ/mol, see figure 9) [35]. Because of this large enthalpy of mixing and the rather low temperature (713K, low configuration entropy), very small mutual solubility is expected (i.e. small Cu amount in fcc Cr-rich precipitates). Let us compute an order of magnitude of the solubility of Cu in Cr precipitates. The parabolic shape of $\Delta H_m$ (~X (1-X), see figure 9 and equation 2) suggests that the CuCr system may be modelled as a regular solid solution. Ignoring the presence of Fe, the solubility of Cu in the fcc Cr phase can thus be expressed as:

$$X^{fcc}_\beta = \exp(-\Omega_{fcc}/RT) \qquad (1)$$

$\Omega_{fcc}$ may be derived from the enthalpy curve ($H^{fcc}$, figure 9) as following:

$$\Delta H^{fcc} = H^{fcc}_{Cu} + X(H^{fcc}_{Cr} - H^{fcc}_{Cu}) + \Omega X(1-X) \qquad (2)$$

Where X is the Cr concentration (see Fig.9). $H^{fcc}_{Cr} - H^{fcc}_{Cu}$ was found close to 7 kJ/mol (figure 9). $\Omega$ has been derived from the first derivative of $H^{fcc}$. This latter is null when:

$$(H^{fcc}_{Cr} - H^{fcc}_{Cu}) + \Omega(1-2X) = 0 \qquad (3)$$

$\Delta H^{fcc}$ exhibits a derivative equal to zero for X = 0.55 (fig. 9) so that $\Omega_{cfc} = 7\ 10^4$ J/mol and $X_\alpha^{fcc} = 7\ 10^{-6}$ at 713K. A similar fit of $H^{bcc}$ leads to $\Omega_{cc} = 10^5$ J/mol and $X_\alpha^{bcc} = 4\ 10^{-8}$ at 713K ($H^{bcc}_{Cu} - H^{bcc}_{Cr} = -4$ kJ/mol and $dH^{bcc}/dX = 0$ pour X = 0.48, see Fig. 9)

In contrast to these predictions (almost pure Cr-precipitates), precipitates are observed with 50% of Cr. This is clearly in disagreement with the well known classical nucleation theory that states that nuclei have the "equilibrium" composition (here that of the metastable phase, i.e. 100% of Cr). Diffusion arguments as well as thermodynamics have both to be considered in order to account for this non-classical nucleation behaviour.

Let us first compute the order of magnitudes of quantities (driving force, elastic energy, interfacial energy) that come into competition in the classical nucleation theory. In the following, we shall compare the classical nucleation barrier (i.e. precipitate composition ~ 100 at.% Cr) to that related to non-classical nucleation (i.e. precipitate composition ~ 50 at.% Cr). The driving force for classical nucleation in dilute alloys reads [36, 37]:

$$\Delta g_n = (RT/V_\beta) X_\beta \ln (X_0/X_\alpha) \qquad (4)$$

with $X_0$ the atomic fraction of solute (Cr) in Cu, $X_\alpha$ the coherent solubility of Cr in the fcc copper phase, $X_\beta$ the Cr concentration in the fcc or bcc chromium phase (close to 100 at.% in theory) and $V_\beta$ the molar volume of the chromium phase. We have assumed this latter to be the same for both bcc and fcc structures of Cr phase.

It must be kept in mind that for a regular solution, $X_\beta = 1-X_\alpha$ (G(X) is symmetrical). The coherent solubility limit of Cr in Cu is $X_\alpha \ll 1$ so that $X_\beta \sim 1$, leading to $\Delta g_n^{fcc} = 6 \cdot 10^9$ J/m$^3$ for the nucleation of fcc Cr and to $\Delta g_n^{bcc} = 10^{10}$ J/m$^3$ for the nucleation of bcc Cr (with $X_0 = 0.7\%$). Let us first compare the "classical" driving force to both the elastic energy resulting from misfits and to the interfacial energy. The classical nucleation barrier will then be computed and compared to the non-classical barrier.

The elastic energy of a coherent fcc precipitate in the fcc copper matrix may be estimated using the classical theory for a coherent inclusion as described by Eshelby [38]:

$$\Delta g_e = (E/(1-\nu)) (\delta a/a)^2 (X_\beta - X_\alpha)^2 \qquad (5)$$

Where E and $\nu$ are respectively the elastic modulus (E = 130 GPa) and the Poisson ratio ($\nu=0,34$) of copper and $\delta a/a$ the misfit between pure fcc-Cu and fcc-Cr phases (1.9% [39]). This leads to $\Delta g_e = 7.4 \cdot 10^7$ J/m$^3$ ($X_\beta = 100\%$), two orders of magnitude lower than the driving force $\Delta g_n$. The elastic energy is negligible compared to the driving force for pure nuclei. It is important to keep in mind that because of the dependence of both the driving force and elastic energy on $X_\beta$, this may be not the case for embryos containing less chromium. This will be discussed latter.

Bcc precipitates exhibit a NW or KS orientation relationship that minimizes the elastic energy, thus for the sake of simplicity, this elastic energy for bcc precipiates will be neglected in the following estimations.

In a first approach, the coherent interfacial energy γ is due to unlike first nearest pair interactions and may be written as proportional to the concentration gradient squared:

$$\gamma = \varepsilon \left((X_\beta - X_\alpha)/a\right)^2 \quad (6)$$

where a is the lattice parameter (abrupt interfaces were here considered) and ε the ordering effective energy that can be written as follow:

$$\varepsilon = \Omega / N\, Z \quad (7)$$

Where N is the Avogadro constant (N=6.02 $10^{23}$ mol$^{-1}$) and with Z the coordinance (Z=12 for fcc and 8 for bcc). Again γ depends on the composition of the nucleus. For pure nuclei, γ = 86 mJ/m$^2$ (coherent $Cr_{fcc}/Cu_{fcc}$ interface). One should note that it is much smaller than the $Cr_{bcc}/Cu_{fcc}$ interfacial energy reported in the literature for non-coherent Cr-Cu fcc/bcc interfaces (625 mJ/m$^2$ [40]). The nucleation barrier (ΔG*) for fcc precipitates will therefore be lower than that of bcc. This latter writes:

$$\Delta G^* = \frac{16\pi}{3} \frac{\gamma^3}{(\Delta g_n + \Delta g_{el})^2} \quad (8)$$

For an fcc nucleus, this leads to $\Delta G^*_{fcc} \sim 3\ 10^{-22}$ J, thus two order of magnitude smaller than for a bcc nucleus $\Delta G^*_{bcc} \sim 4\ 10^{-20}$ J. This clearly confirms that the nucleation of fcc Cr-enriched nuclei is much easier than that of non-coherent bcc nuclei.

Let us now consider non-classical nucleation [41-45]. As shown in figure 10, lower Cr content in embryos significantly reduces the driving force for the nucleation. However, nuclei containing less Cr will give rise to lower interfacial energy because γ is proportional to the square function of the concentration gradient ($\nabla X$) (equation 6). Lower solute (Cr) content in embryos also reduces the elastic energy (equation 5) that also acts as a barrier against nucleation. The expected composition of critical nuclei will be therefore that which minimizes the nucleation barrier (saddle point of the barrier). This situation is schematically represented in Fig. 10. Considering again a regular solid solution, it is easy to show that the non-classical driving force for the nucleation of precipitates that have a solute content $X_\beta$ writes as [44]:

$$(\Delta g_n)^{NC} = \frac{RT}{V_\beta}\left[\frac{2}{\eta}(X_\beta - X_0)^2 - \left(X_\beta \ln\left(\frac{X_\beta}{X_0}\right) + (1-X_\beta)\ln\left(\frac{1-X_\beta}{1-X_0}\right)\right)\right] \quad (10)$$

with $\eta = T/T_c$ and $T_c$ the critical temperature, $T_c = Z\varepsilon/2k$ ($T_c = 4870K$ for $\varepsilon = 70$ meV and $Z=12$ (fcc), $\eta = 0.15$ for T = 713K).

The driving force yields $(\Delta g_n^{fcc})^{NC} = 1.35\ 10^9$ J/m$^3$, a value that is 5 times smaller than the classical value ($\Delta g_n^{fcc} = 6\ 10^9$ J/m$^3$). The elastic energy, proportional to $(X_\beta - X_\alpha)^2$, is close to one fourth of the previous value for 50% of Cr instead of 100% : $\Delta g_e = 1.85\ 10^7$ J/m$^3$. This value is two orders of magnitude smaller than the driving force. Elastic energy due to misfit cannot therefore account for the lower Cr content in nuclei. In the framework of classical nucleation theory, a lower solute content in nuclei is expected to favour a lower misfit, thus reducing the elastic energy and the nucleation barrier. However, if we instead consider non-classical theory of nucleation, increasing the elastic energy unexpectedly increases the solute concentration in nuclei [45]. In addition, the above estimations show that the elastic contribution can be ignored and anyway this would not lead to an underestimation of Cr content in precipitates.

In contrast, the influence of the interfacial energy is considerable. For $X_\beta = 0.5$, the interfacial energy is $(0.5)^2$ smaller ($\gamma = 21.5$ mJ/m$^2$, equation (6)) so that this contribution in the barrier expression (equation (8)) is considerably smaller $(1/2^2)^3 = 1/64$ times smaller (compared to the decrease of the driving force $\Delta g_n^{fcc} \sim (1/5)^2$. This leads to a nucleation barrier $\Delta G^*_{NC}$ of only $9\ 10^{-23}$ J for fcc embryos containing 50at.%Cr. This value of the barrier height is close to ten times smaller than the classical value that we have previously computed for pure Cr embryos ($3\ 10^{-22}$ J). These thermodynamical considerations based on the non-classical theory of nucleation indicate that it is much easier to nucleate precipitates containing 50at% of Cr compared to pure Cr precipitates. The related critical radius can be computed as following:

$$r^* = 2\gamma / \Delta g_n \qquad (11)$$

This leads to $r^*_{fcc} = 2.9 \; 10^{-11}$ m for non-classical fcc nuclei and $r^*_{bcc} = 1.2 \; 10^{-10}$ m for bcc nuclei. These latter values are unfortunately smaller than the atomic size. This clearly raises the problem of the validity of nucleation theories for such large supersaturations and driving forces.

In summary, the equilibrium structure of Cr precipitate is bcc, and their equilibrium Cr content is close to 100%. However, the nucleation barrier for fcc precipitates is lower than that of bcc precipitates. For low Cr content in nuclei, the nucleation barrier is even lower. For 5h and 10h at 713K, both bcc and fcc precipitates were observed. Because of their much smaller nucleation barrier, dilute fcc precipitates first appear. bcc precipitates then appear at the expense of fcc nuclei. The proportion of each is controlled by the respective values of both the nucleation rate and the nucleation latent times.

The nucleation pathway of incoherent phases is a generic issue that is common for many systems where cascades of metastability are observed (AlCu, AlMgSi, FeCu…). Fcc nuclei being not the stable form of the Cr phase, their "stability" decreases in proportion as they grow so that they finally disappear at the expense of the stable bcc Cr-rich phase. The driving force is of course the misfit elastic energy of coherent fcc precipitates that dramatically increases with size while the interfacial energy becomes less important in the overall energy balance. Then, three mechanisms might be considered:

i/ Heterogeneous precipitation of the bcc phase on fcc nuclei. The driving force for such a heterogeneous nucleation is not the reduction of interfacial energy as interfaces are coherent. Instead it may originate from the presence of the large amount of Cr present in fcc nuclei. Nucleation is thus easier and faster. However, APT data never shown coagulated particles, thus this mechanism can be ruled out.

ii/ Dissolution of fcc nuclei, leading to an increase of the supersaturation and the subsequent homogeneous nucleation of the bcc phase that has a larger latent time for its nucleation. The calculation conducted above shows that the driving force remains very large even for 50% of Cr. This mechanism is therefore quite plausible and also consistent with our experiments that clearly show that bcc and fcc precipitates co-exist.

iii/ Transformation of the whole structure of nuclei from fcc to bcc during the course of their growth. This process involves the massive transformation of the precipitate structure as observed in diluted Fe-Cu alloys.

Up to now, the influence of iron was not considered. This element was found to partition preferentially to precipitates (fig. 7(a)). Even if this third element is present with a low content, it may have a drastic influence on the low Cr content observed in precipitates. Because of the low miscibility between Fe and Cr (large miscibility gap), the presence of Fe in the core observed in the early stages (fig 7(a)) may reduce significantly the amount of Cr in particles especially in the early stages. This interpretation is reinforced by the fact that the Cu concentration in precipitates decreases (i.e. the Cr content decreases) when iron is rejected at the matrix/precipitate interface during their growth (fig. 7(c)).

Kinetics effects have also to be considered. The diffusion coefficient of Cu in bcc Cr at 713K (4.1 $10^{-17}$cm$^2$ s$^{-1}$ [46]) is two order of magnitude smaller than that of Cr in fcc Cu (1.5 $10^{-15}$cm$^2$ s$^{-1}$ [47]). Thus, compared to the rate at which Cr atoms migrate and agglomerate to form nuclei, the rejection of Cu atoms from bcc Cr-rich precipitates will take more time and precipitates are likely to grow and reach larger sizes (a few nm) before reaching their equilibrium composition (i.e. close to 100% Cr).

The presence of Fe at the interface is also intriguing and deserves to be discussed further. Fe is mainly located in the core of precipitates in the early stages of aging. A rejection of this element to Cu/Cr interfaces is observed during coarsening (Fig. 7). This thin Fe-rich layer formed at the interface leads to the formation of two new interfaces (Cu/Fe and Fe/Cr) and the elimination of Cu/Cr interfaces. This raises the question of whether this transformation is favorable energetically. Cu/Fe interfacial energy $\gamma_{Cu/Fe}$ is in the range 0.121 to 0.318 J m$^{-2}$ [48] whereas $\gamma_{Fe/Cr}$ is reported to lie between 0.047 to 0.213 J m$^{-2}$ [49]. These interfacial energies are much smaller than that of Cu$_{fcc}$/Cr$_{bcc}$ interfaces (0.625 J/m$^2$ [40]). Even if the core-shell structure observed for such precipitates (Fe/Cr/Fe) leads to the formation of two interfaces instead of one, segregation of Fe seems quite favorable even when adding interfacial energies.

## 5. Conclusions

The precipitation of Cr in fcc Cu has been investigated using TEM and APT to characterize nanoscaled precipitates that nucleate during the early stage of aging. Three kinds of precipitate were observed: spheres (S), ellipsoids (E) and plates (P). Spheres are the smallest in size, they contain between 30 and 60 at.% Cr and are attributed to fcc coherent precipitates. Ellipsoids are somewhat bigger in volume than spheres, they exhibit a higher Cr content (between 60 and 80 at.%), and are attributed to bcc precipitates with a NW-OR. Plate shaped precipitates are the biggest in size, they exhibit the highest Cr content (in a range of 85 to 100 at.%) but are almost never pure as expected from the phase diagram ; they are attributed to bcc precipitates with a KS-OR.

The observation of fcc Cr rich precipitate was rather surprising because only bcc precipitates were expected from the phase diagram. However, using the classical nucleation theory, it was found that the nucleation barrier for fcc precipitates is much lower (2 order of magnitude) than for bcc precipitates. However, the classical theory could not account for the rather high amount of Cu measured by APT in the fcc precipitates. Thus, the non-classical nucleation theory was applied taking into account both the elastic and the interfacial energies. It was found that a precipitate composition of about 45at.%Cr was minimizing the nucleation barrier. Then, during the coarsening stage, precipitates transform into bcc and progressively increase their Cr content.

|  | $X_m$ (at. % Cr) | $N_v$ ($10^{23}$ m$^{-3}$) | $F_v$ (%) | Hv (kg/mm$^2$) | σ % IACS |
|---|---|---|---|---|---|
| As quenched | 0.78 ± 0.09 | --- | --- | 55 ± 5 | 31 ± 6 |
| 5 h | 0.14 ± 0.003 | 11.6 ± 0.9 | 1.3 ± 0.2 | 145 ± 8 | 61 ± 6 |
| 10 h | 0.022 ± 0.008 | 2.7 ± 0.4 | 0.86 ± 0.1 | 155 ± 8 | 76 ± 10 |

**Table 1**

Summary of the experimental data collected for the copper alloy in the as quenched state, after 5h and 10h of aging at 713K. $X_m$ (Cr concentration in the fcc copper matrix), $N_v$ (precipitate density) and $F_v$ (volume fraction of precipitates) are estimated from APT data. Hv is the Vickers microhardness and σ is the electrical conductivity given as a fraction of the International Annealed Copper Standard (IACS) 58 $10^6$ S/m.

**Acknowledgements**

The authors would like to thank Dr. C. Genevois for assistance in TEM. Dr Eiji Okunishi from JEOL Ltd is also gratefully acknowledged by the authors for the EDS and EELS STEM analysis.

# Figures

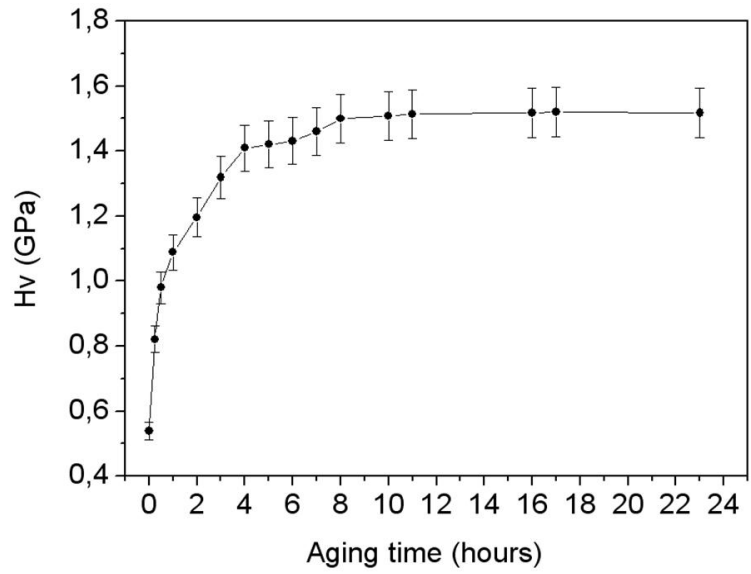

**Figure 1:** Hardness of the Cu alloy as a function of the aging time at 713 K.

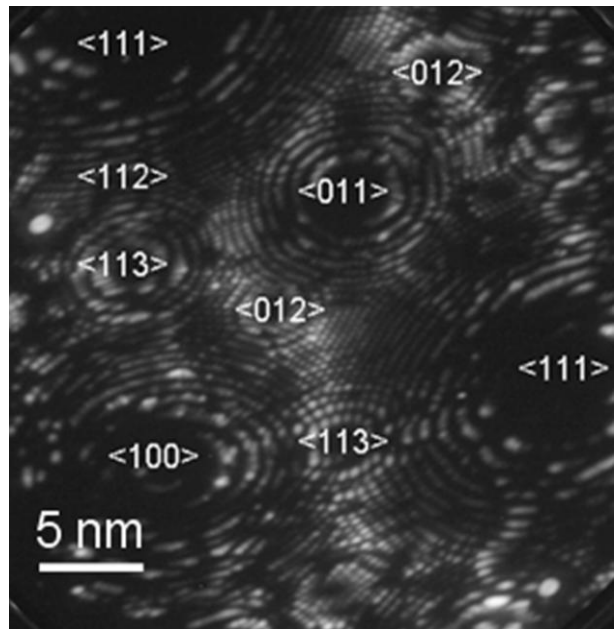

**Figure 2:** Field ion microscopy image of the Cu alloy aged during 10h at 713K. There is no visible contrast between the copper matrix and the chromium precipitates indicating that these two phases have similar evaporation fields.

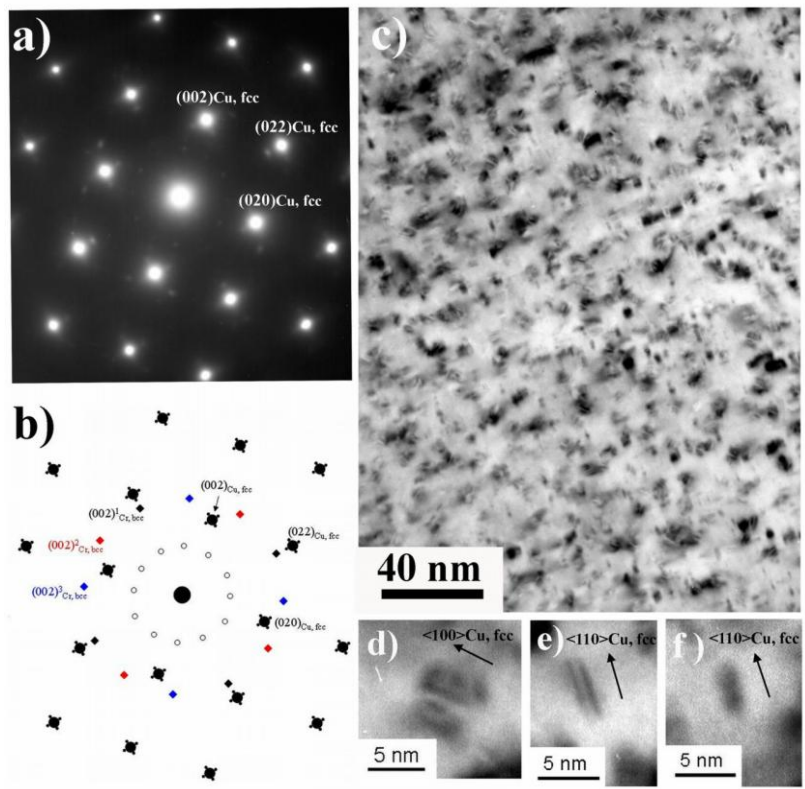

**Figure 3:** TEM images of the Cu alloy aged 10h at 713K: Selected area electron diffraction pattern collected in the (002) fcc Cu zone axis (a), indexation of the full pattern – see text for details (b), bright field image obtained with a specimen orientation slightly tilted from the (002) fcc Cu zone axis (c), zooms on nanoscaled contrasts attributed to three different kind of precipitates (d), (e) and (f) – see text for details.

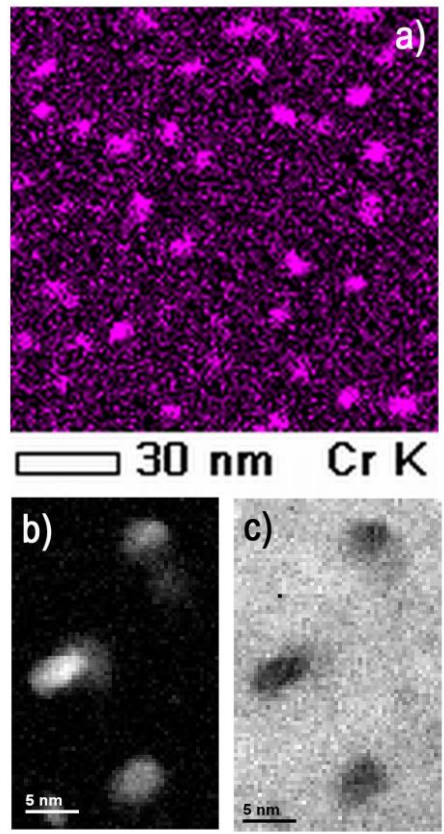

**Figure 4:** Analytical STEM images of the Cu alloy aged 10h at 713K: EDX Cr map (Cr k-alpha signal) showing the high density of nanoscaled Cr rich precipitates (a); EELS Cr (b) and Cu (c) maps showing three Cr rich precipitates. The biggest is a 5nm long ellipsoid

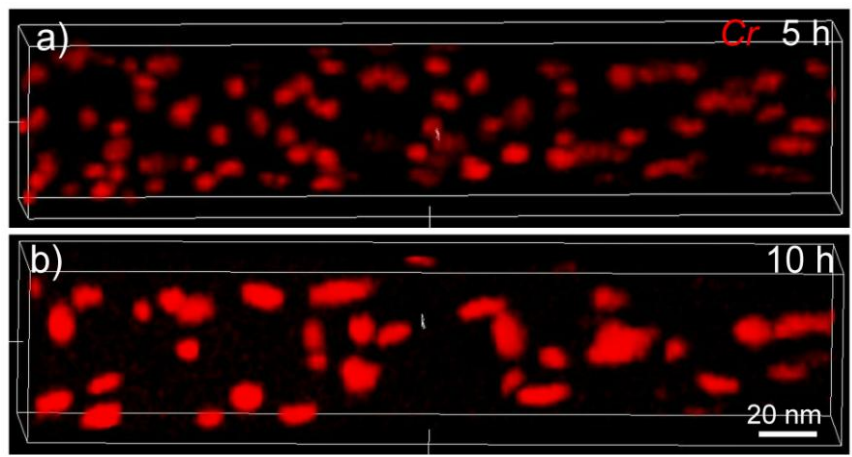

**Figure 5:** APT data showing 3D chromium density maps (30x30x140 nm$^3$) in the Cu alloy aged 5h (a) and 10h (b) at 713K.

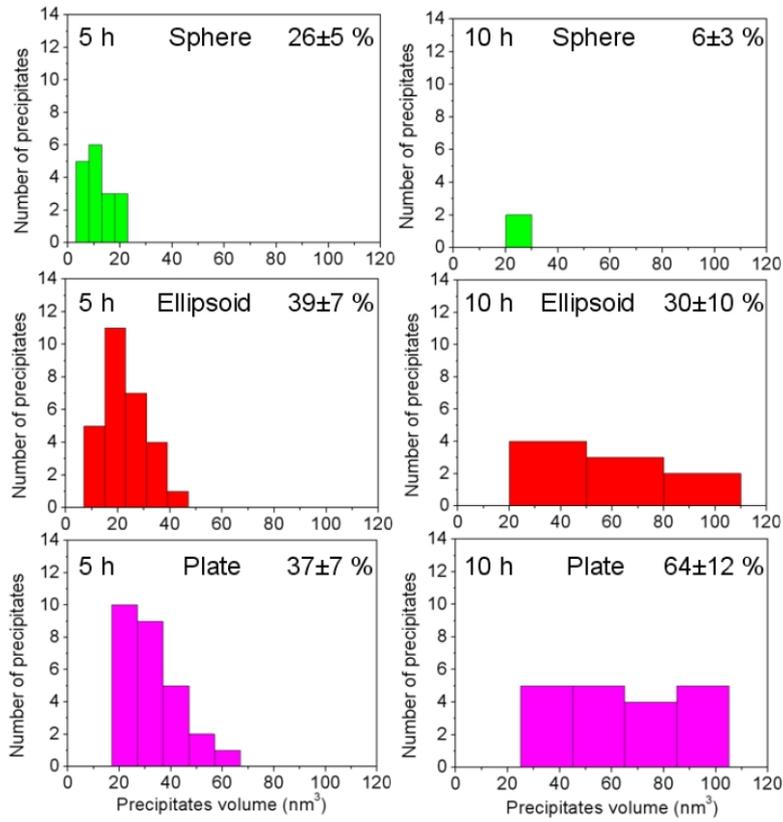

**Figure 6:** Size distribution of the three different kind of precipitates analyzed by APT (Spheres "S": (a) and (d) ; Ellipsoids "E": (b) and (e), Plates "P": (c) and (f)) after aging at 713 K during 5h ((a), (b) and (c)) and 10h ((d), (e), (f)). The proportion of each kind of precipitate for a given aging time is indicated in the top right corner of the histogram.

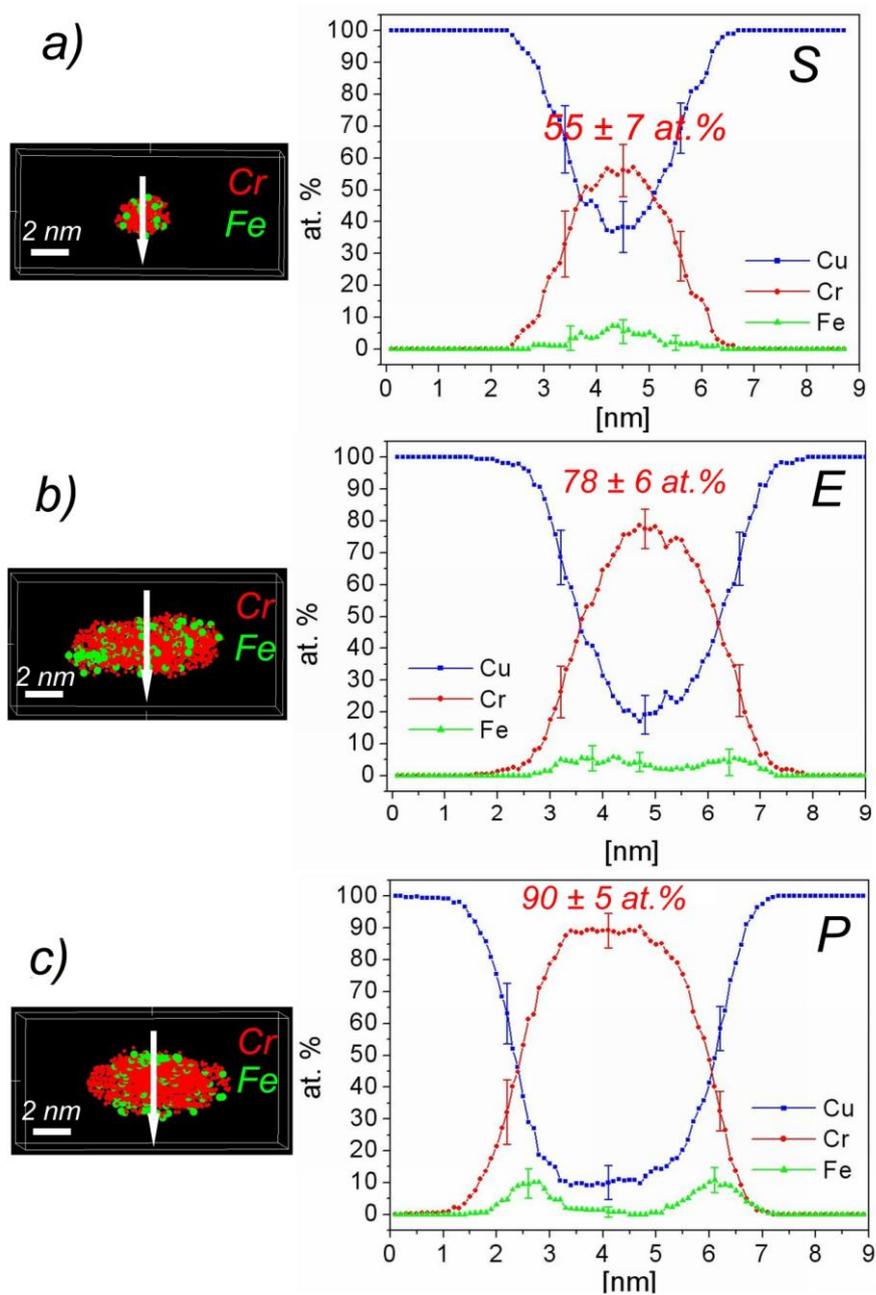

**Figure 7:** 3D reconstruction of typical precipitates (Spheres (a), Ellipsoid (b) and Plate (c)) and corresponding concentration profiles computed across (the arrows indicate the location of the profiles, sampling volume thickness: 1 nm). In the reconstructed volumes only Cr (red) and Fe (green) atoms are displayed.

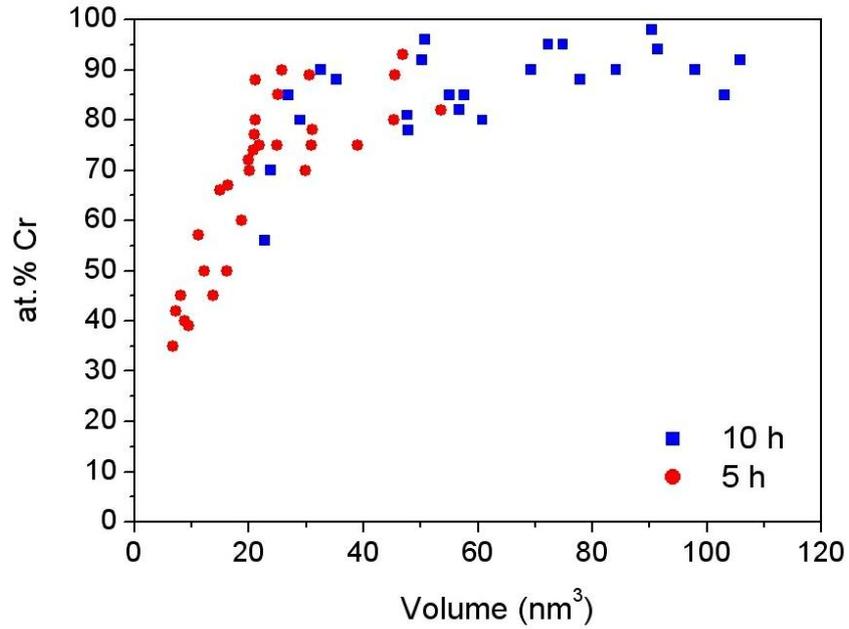

**Figure 8:** Chromium concentration of the precipitates plotted as a function of their size for the alloy aged 5h (red circles) and 10h (blue squares) at 713K. Obviously the bigger the precipitate the higher is the Cr content.

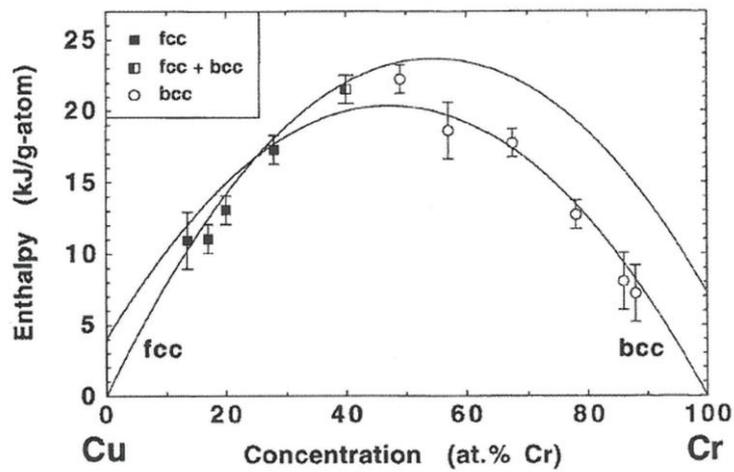

**Figure 9:** Formation enthalpy of non equilibrium super saturated solid solution in the Cu-Cr system (from [35])

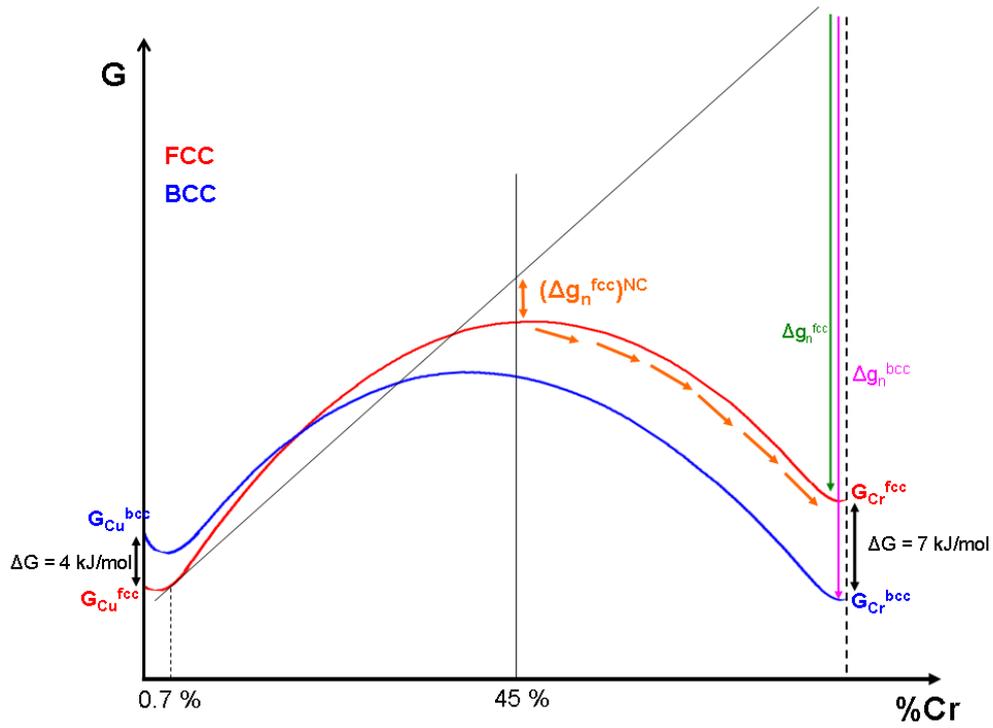

**Figure 10:** Schematic representation of the free enthalpy curves of the fcc and bcc phases of the Cu-Cr system (established from the data of Fig.9), showing the driving force for the nucleation of the fcc ($\Delta g_n^{fcc}$) and bcc ($\Delta g_n^{bcc}$) Cr rich phases, and also the driving force considering the non-classical theory of germination $(\Delta g_n^{fcc})^{NC}$.